\documentclass[aps,prx,twocolumn,reprint, superscriptaddress,nofootinbib,notitlepage,longbibliography]{revtex4-1}
\usepackage{extpfeil}

\usepackage{times} 

\usepackage{booktabs}
\usepackage{graphicx}
\usepackage{amsmath}
\usepackage{amstext}
\usepackage{amssymb}
\usepackage{romannum}
\usepackage[colorlinks,citecolor=blue]{hyperref}
\usepackage{graphicx}
\usepackage{amsmath}
\usepackage{amstext}
\usepackage{amssymb}
\usepackage{amsfonts}
\usepackage{longtable,booktabs}
\usepackage{hyperref}\usepackage{url}
\usepackage{subfigure}%
\usepackage{dsfont}
\usepackage{datetime}

\usepackage{amsbsy}
\usepackage{dcolumn}
\usepackage{amsthm}
\usepackage{bm}
\usepackage{esint}
\usepackage{multirow}
\usepackage{hyperref}
\hypersetup{
    colorlinks=true,
    linkcolor=magenta,
    filecolor=cyan,
    urlcolor=blue,
}
\usepackage{booktabs}

\usepackage{ctable}
\usepackage{cleveref}
\usepackage{comment}
\usepackage{mathrsfs}
\usepackage{amsfonts}
\usepackage{amsbsy}
\usepackage{dcolumn}
\usepackage{bm}
\usepackage{multirow}
\usepackage{color}
\newcounter{defcounter}
\usepackage{amsthm}

\def\Z{\mathbb{Z}}

  \usepackage{extarrows}
 \usepackage{graphicx}
    \usepackage{amssymb}
    \usepackage{amsthm}
    \usepackage{mathtools}
    \usepackage{adjustbox}

\usepackage{hyperref}
\hypersetup{
    colorlinks=true,
    linkcolor=blue,
    filecolor=blue,
    urlcolor=blue,
}
\def\Z{\mathbb{Z}}

\def\c{\chi}
\def\p{\partial}
\def\m{\mathcal}
\def\D{\Delta}
\def\d{\delta}
\def\T{\Theta}

\def\o{\omega}
\def\t{\theta}

\def\O{\Omega}

\begin{document}
\title{Generalized Wen-Zee    Terms }

\pagenumbering{arabic}

\author{Bo Han}
\affiliation{Department of Physics and Institute for Condensed Matter Theory, University of Illinois at Urbana-Champaign, Illinois 61801, USA}

\author{Huajia Wang}
\affiliation{Kavli Institute for Theoretical Physics, University of California, Santa Barbara, California, 93106, USA}
\affiliation{Department of Physics and Institute for Condensed Matter Theory, University of Illinois at Urbana-Champaign, Illinois 61801, USA}

\author{Peng Ye}
\email{yepeng5@mail.sysu.edu.cn; yphysics@illinois.edu}
\affiliation{School of Physics, Sun Yat-sen University, Guangzhou, 510275, China}
\affiliation{Department of Physics and Institute for Condensed Matter Theory, University of Illinois at Urbana-Champaign, Illinois 61801, USA}

\begin{abstract}
Motivated by symmetry-protected topological phases (SPTs) with both   spatial symmetry (e.g., lattice rotation) and internal symmetry (e.g., spin rotation), we propose a class of exotic topological terms,  which   generalize the well-known Wen-Zee topological terms  of  quantum Hall systems [X.-G.~Wen and A.~Zee, Phys.~Rev.~Lett.~$\textbf{69}$, 953 (1992)].  These generalized Wen-Zee terms   are expressed as   wedge product of spin connection   and usual  gauge fields (1-form or higher) in various dimensions.   In order to probe   SPT orders,  we  externally insert  ``symmetry twists'' like    domain walls of discrete internal symmetry and disclinations that are geometric defects with nontrivial Riemann curvature.  Then,   generalized Wen-Zee terms  simply tells us how SPTs respond to those symmetry twists. Classifying these exotic  topological    terms  thus leads to a complete classification and characterization of SPTs within the present framework. We  also propose SPT    low-energy field theories, from which   generalized Wen-Zee terms   are  deduced  as topological response actions.  Following   the Abstract of Wen-Zee paper,  our work  enriches  alternative  possibilities of condensed-matter realization of unification  of electromagnetism and ``gravity''.
 \end{abstract}


\maketitle


\section{Introduction}

\textit{SPTs with   internal symmetries}.--- Topological insulators and topological superconductors  have been a large platform for both theoretical and experimental studies in condensed matter community since their discovery.~\cite{Qi2011TIreview, Hasan2010TIreview} Recently they have been recognized to belong to a large class of symmetry-protected topological (SPT) phases. SPT phases are topological phases that can be adiabatically connected to the trivial phase if the symmetry is broken explicitly or spontaneously. Nevertheless, if the symmetry persists, there are several nontrivial  properties that are protected by symmetry, such as nontrivial edge states on a manifold with boundaries.   Stimulating efforts towards classification and characterization of SPT orders have been made, from free-fermion phases (for a recent review, see Ref.~\onlinecite{Chiu2016RMP}) to   boson / spin systems \cite{Chen2013CGLWbSPT}. Especially, internal symmetries, which \textit{simultaneously} act on internal space (e.g., spin space of spin models) of each lattice site leaving  spatial coordinates unaltered, have been systematically studied through different approaches in bosonic systems, such as group cohomology \cite{Chen2013CGLWbSPT}, cobordism groups \cite{Kapustin2014Cobordism, Kapustin2015fCobordism}, non-linear sigma models \cite{Bi2015NLSMSPT,YouYou2016bSPT}, topological field theories \cite{Lu2012CSSPT,bti2,Gu2016SPTBF,Ye20163dbSPT,2018arXiv180105416W}, conformal field theories  \cite{Ryu2012SPTMI,Hsieh2014SPT2dorbifold,Hsieh2016SPT3dorbifold,Han2017BCFTSPT}, decoration picture \cite{Chen:2014aa}, topological response / gauged theory \cite{PhysRevLett.112.141602,Hung_Wen_gauge,PhysRevD.88.045013,bti6,lapa17,Wang2015GaugeGravitySPT}, and projective / parton construction \cite{YW12,LL1263,Ye14b,YW13a,ye16a}, braiding statistics approach \cite{levin_gu_12,wang_levin1,2016arXiv161209298P,ye17b} in different spatial dimensions.  

\textit{SPTs with spatial symmetries}.--- In comparison to internal symmetries, spatial symmetries, however, have not been systematically explored until recent years. Spatial symmetries act on either lattice geometry in lattice models or continuum spatial coordinates in continuum topological media.  Such symmetries were  introduced in the study of  topological phases in Ref.~\onlinecite{Fu2011TCI} via free fermion systems and later enriched the ten-fold classifications. \cite{Chiu2016RMP} For the past few years, there have been   several progresses on the classification of three dimensional SPT phases protected by point group symmetries with reflection symmetry, called point group SPT (pgSPT). \cite{Song2017pgSPTPRX,Huang2017cSPTPRB,Huang2017SFTpgSPT,SongHuangQiFangHermele2018topologicalcrystal} Their central idea follows from Ref.~\onlinecite{Isobe2015IntTCI}, namely, to reduce the bulk SPT phases to lower dimensional SPTs on the reflection plane via local unitary transformations, and study the classification of pgSPT phases on the reflection plane, where the $\mathbb{Z}^P_2$ reflection symmetry becomes an internal $\mathbb{Z}_2$ symmetry. The authors of Ref.~\onlinecite{Song2017pgSPTPRX,Huang2017cSPTPRB,Huang2017SFTpgSPT}
 study 3D pgSPT phases for both bosonic and fermionic systems in a case-by-case manner and obtain the classifications by studying lower dimensional building blocks, from both algebraic and field theoretic approaches. Their results agree with those in Ref.~\onlinecite{Thorngren2016Gaugingspatial}, where a more mathematical approach was developed. In addition to dimensional reduction, Ref.~\onlinecite{Jiang2017SPT} utilizes the construction of the generic tensor-network wave functions and group cohomologies for the classification of SPT phases including crystalline symmetries.

\textit{Mixture  leads to more}.--- 
In this paper, we  focus on a new class of SPT phases  in boson / spin systems with   mixture of spatial and  internal symmetries.  
 Naively, one can achieve such SPTs through a simple stacking of  two constituents: an SPT with only spatial symmetry and an SPT with only internal symmetry. Nevertheless, such a kind of construction is of no interests since the spatial and internal symmetries do not truly  talk to each other.  What we really want to explore are those SPT phases that are always  adiabatically connected to the trivial phase once any part of  symmetry is broken. In other words, in order to make SPT nontrivial, neither spatial nor internal symmetries are allowed to be broken.  In this paper, we  perform a topological quantum field theory (TQFT) analysis in both $2+1$d and $3+1$d and   construct a class of new topological terms, namely, \emph{generalized Wen-Zee terms} \cite{Wen1992WZ}\footnote{For a brief review, see  Appendix \ref{appendix_review_WZ}.}, in order to characterize and classify SPT orders jointly protected by spatial and internal symmetries.  As an initial attempt, throughout the  paper we only consider the simplest combinations (direct product) of symmetry groups: the  rotational symmetry $C_{N_0}$ or SO(2) and internal symmetry $\Z_{N_1}\times\Z_{N_2}\times\cdots$ or $\text{U}(1)$.

Interestingly, geometric defects play a  critical role in probing spatial rotation symmetry. It is well known that  in the theory of geometric defects in real materials \cite{Katanaev19923dGravityDefect,Nakaharabook},  there are  two typical types of defects, namely, dislocations and disclinations, that can be interpreted in terms of  Riemann-Cartan geometric quantities: torsion and curvature fluxes respectively. The former can be used to probe translational symmetry while the latter can be used to probe  rotational symmetry. Therefore, in order to probe the SPT order of interests, we properly  insert   disclinations of rotational symmetry and  symmetry flux lines (or domain walls) of the internal symmetry.

In Sec.~\ref{section_response}, we find that the resulting response theory is governed by a new type of topological terms (see Table~\ref{tab:Mixed1}), which are generalization of the celebrated Wen-Zee term \cite{Wen1992WZ} as mentioned above. These topological terms mix the  spin connection ($\omega$) and the usual gauge fields (1-form fields $A^1,\,A^2$ and 2-form field $B\cdots$). Since the presence of spin connection sources disclination lines in the bulk, the  topological terms give rise to a topological interaction between gravitational  and electromagnetic degrees of freedom,  implying  an exotic entanglement between general relativity and electromagnetism.  By enforcing the gauge invariance under large gauge transformations, the coefficients of topological terms are quantized and periodically identified, which can be used to characterize and classify the  underlying SPTs.   We note that, only 2+1 and 3+1d are constructed explicitly. In fact, generalized Wen-Zee terms can be simply constructed in all other dimensions in a similar fashion.

In Sec.~\ref{section_bulk_field} we study   bulk low-energy field theories of SPTs. We explicitly show that integrating out  bulk degrees of freedom (i.e., dynamical gauge fields) leads to the generalized Wen-Zee terms discussed in Sec.~\ref{section_response}. Several technical details as well as other relevant   discussions are present in Appendices.   We conclude this paper in Sec.~\ref{sec:Discussion} with discussions and future directions of our work.  

\section{Generalized Wen-Zee terms as topological response actions}
\label{section_response}
We introduce some preliminary geometries in Sec.~\ref{sec:Preliminary}. Specifically, we introduce topological defects in solids described by Riemann-Cartan geometry. We will focus on the curvature part because it is related to the SPT phases that will be discussed in this paper. The torsion part is relegated in Appendix~\ref{append:dislocation}. Readers who are familiar with the gauge theory interpretation of Riemann-Cartan geometry in defects of solids may skip this part. Sec.~\ref{sec:ClassResponse} presents general principles for constructing topological terms (i.e., generalized Wen-Zee terms) and classification principles, from which the classifications of different topological theories are obtained. Sec.~\ref{sec:Interpretation} provides their physical interpretations. Some topological terms in (2+1)d and (3+1)d  are analyzed in details.
\subsection{Preliminaries: Geometric defects and Laughlin argument}
 
\label{sec:Preliminary}
 
In (3+1)d spacetime, Poincar\'e invariance is usually explicitly broken in materials. What we can physically probe is only a subgroup of the whole Poincar\'e group, namely, $SO(3)$ rotation group and the translation group. Real materials are not perfect. In the bulk or on the surface of the material, both of the two subgroups can be broken, leading to different kinds of defects. Translation-breaking defects are called {\it dislocations}, while rotation-breaking defects are called {\it disclinations}. In terms of geometry, deformations in materials can be described by Riemann-Cartan geometry, which can also be interpreted as gauge theories. Dislocations are reflected by the nonvanishing torsion tensor $T^{ij} \neq 0$ while disclinations are reflected by the nonvanishing curvature tensor $R^{ij} \neq 0$~\cite{Katanaev19923dGravityDefect,Landau1986elasticity,Hehl2007Cartan},  
which are defined below. In this paper, the disclination will be used as a background field to probe nontrivial SPT phases. 
 
In Ref.~\onlinecite{Hughes2013Torsion}, the dislocation has been regarded as the torsion flux. From Laughlin argument, an insertion of dislocation would result in the pumping of momentum, which is reviewed in Appendix~\ref{append:dislocation}. Here we would like to make an analogy for disclinations. That is, the disclination can be regarded as the curvature flux; an insertion of disclination would result in the pumping of angular momentum or spin. Since normal gauge fields and their gauge fluxes can be used as background fields to probe and classify SPT phases protected by internal symmetries,~\cite{Ye20163dbSPT} it is reasonable to utilize the $SO(2)$ spin connection and its curvature flux to probe and classify SPT phases protected by spatial symmetries. 

Let $u^i(x)$ be the displacement field at point $x \in \mathbb{R}^3$. After some small deformation, the point $x$ is shifted to $y^i(x) = x^i + u^i(x)$. Then from mathematical point of view, under this diffeomorphism transformation, the Euclidean metric changes as \cite{Katanaev19923dGravityDefect} $g_{ij} = \frac{\p x^k}{\p y^i} \frac{\p x^l}{\p y^j} \d_{kl} \cong \d_{ij}-2\epsilon_{ij}$, where $\epsilon_{ij} \equiv \frac{1}{2} (\p_i u_j+\p_j u_i)$ is the strain tensor. Note that here we assumed that the deformation is small. One can check that from this metric, the deformed three manifold has zero torsion and zero curvature. This makes sense because under small deformations, there is no room for defects like dislocations and disclinations, which are described by nonzero torsion and curvature, respectively. In spite of the absence of defects, these small deformations are still physically real and have physical effects. For instance, the elastic waves in these deformed media will not propagate along straight lines because geodesics are curved. 

A disclination in the media is represented by the Frank vector $\mathbf{\T}$. Once we fix a reference direction $n^i_0$ in the media, then any direction $n^i(x)$ can be described by rotational angles $\alpha^{ij} \in SO(3)$ where $\alpha^{ij}= -\alpha^{ji}$. Then we can define $\oint_C dx^m \p_m \alpha^{ij} = \O^{ij}$. 
In terms of the displacement field, we have~$\alpha_{ij} \equiv  \p_i u_j-\p_j u_i$. The Frank vector is defined as $\T_i = \epsilon_{ijk} \O^{jk}$. Here $\alpha^{ij}$ is not a smooth field in the presence of disclinations. Nevertheless we can always require that $\p_m \alpha^{ij} \equiv \o_m^{~ij}$ is a smooth field. Then we have 
$
\O^{ij} \sim \oint_C dx^m \o_m^{\ ij} = \iint_S \frac{1}{2}dx^m dx^n (\p_m \o_n^{\ ij} - \p_n \o_m^{\ ij}),  
$ 
where $\p S = C$. When disclinations exist, $\O^{ij} \neq 0$. Generally, in terms of the curvature tensor $R_{mn}^{\quad ij} \sim \p_m \o_n^{\ ij} - \o_m^{\ ik} \o_{nk}^{\quad j} - (m \leftrightarrow n)$,  we can generically have $\iint_S dx^m dx^n R_{mn}^{\quad ij} = \O^{ij}$. It is reminiscent of the electromagnetism that $ \bm{S} \cdot \bm{B} =  \Phi$, where $\Phi$ is the magnetic flux and $\bm{B}$ is the magnetic field. Now it is easy to interpret curvature as the surface density of the curvature flux, whose strength is represented by the Frank vector.   Laughlin argument tells us that an insertion of the disclination induces angular momentum pumping. In this sense, angular momentum is the {\it charge} if we regard the spin connection as a gauge field. 

\begin{figure}[htbp]
\includegraphics[width=0.47\textwidth]{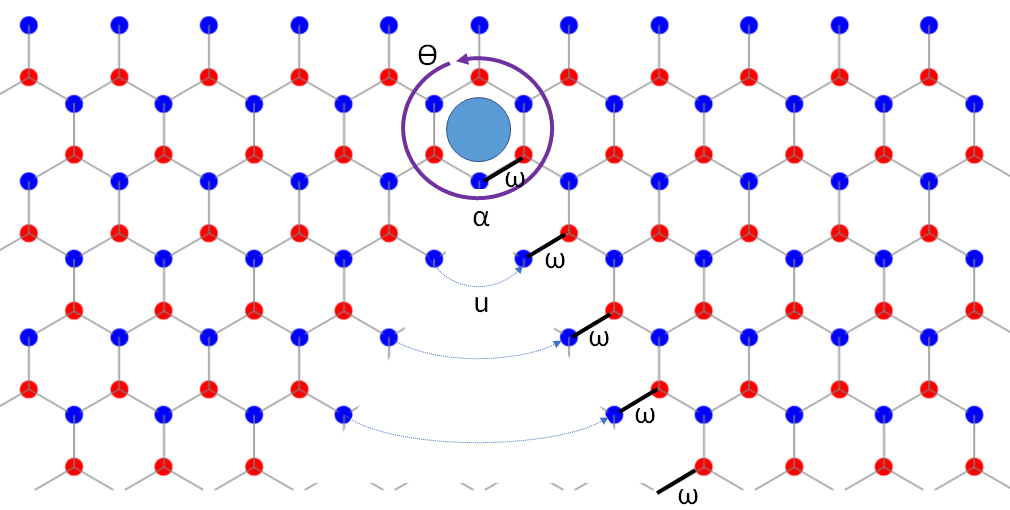}
\caption{(Color online). Schematic diagram for creating a disclination via Volterra process on a 2D lattice on the $xy$ plane. First, part of the lattice is removed and then the two wedges are glued together, with identification of the corresponding lattice sites connected by blue arrows. Here $u$ is the dispacement field and $\alpha$ is the bond angle defined by $\alpha= \epsilon_{ij} \p_i u_j$. $\o = d \alpha$ is the effective U(1) connection defined on the links, introducing a branch cut after the gluing process. Finally, $\T=\oint_C \o$ is the Frank vector along the $z$ direction produced by $\o$, where $C$ is the purple loop enclosing the origin $0$. $\o$ is the spin connection defined on the corresponding links. It is easy to see that $\T$ is the flux of $\o$. The blue shaded region is the regularized disclination flux.}
\label{fig:Volterra}
\end{figure}

Fig.~\ref{fig:Volterra}   shows the creation of a disclination on a 2D lattice via Volterra process and the geometric relation between the displacement field $u^i$, the bond angle $\alpha \equiv \alpha^{xy}$, the spin connection $\o \equiv \o^{xy}$ and the Frank vector $\bm{\T} = (0,0,\T)$ defined above, in the $xy$ plane, for simplicity. Specifically, $\alpha = \epsilon_{ij} \p_i u_j$, $\o = d \alpha$ and $\T = \oint_C \o$. If we restrict to the rotation $SO(2)$ subgroup, then the general Lorentz transformation is reduced to $\o \to \o + d\t $ 
which is isomorphic to a U(1) transformation. Then we can regard $\o$ as a usual U(1) gauge field to discuss SPT classifications, which is analogous to what has been done for internal symmetries~\cite{Ye20163dbSPT}. It is good to understand this gauge field at the level of displacement field $u^i$. At the level of the bond angle and the displacement field, the U(1) gauge transformation $ \o \to \o + d\t$ is equivalent to $\alpha \to \alpha + \theta$ and $u^i \to u^i + v^i$, where $\epsilon_{ij} \p_i v_j = \t$ on the $ij$ plane.

\subsection{Generalized Wen-Zee terms and SPT classification}\label{sec:ClassResponse}
As the SO(2) spin connection can be treated as a U(1) gauge field, we can study the classification of SPT phases protected by both rotational symmetries and internal symmetries from their effective response TQFTs. The main results are summarized in Table \ref{tab:Mixed1} below. 
\begin{table*}[t]
\centering
\begin{tabular}{c|c|c|c|c}
        \specialrule{1.5pt}{1pt}{1pt}
\begin{minipage}{0.1\textwidth}{Spacetime Dimension}\end{minipage} & \begin{minipage}{0.13\textwidth} Spatial Symmetry $G_s$\end{minipage} & \begin{minipage}{0.13\textwidth} Internal Symmetry $G_i$\end{minipage}  &  \begin{minipage}{0.28\textwidth}{{\it Irreducible}  Generalized Wen-Zee terms}\end{minipage} & Angular Momentum (spin) $\m{J}$  \\
        \specialrule{1pt}{1pt}{1pt}
2+1d & $SO(2)$ & $U(1)$ & $\frac{k}{2\pi} \int \o \wedge dA$,  $k \in \mathbb{Z}$ & $\frac{k}{2\pi} \int_{M^2} dA$ \\
 & $C_{N_0}$ & $\mathbb{Z}_{N_1}$ & $\frac{k}{2\pi} \int \o \wedge dA$, $k \in \mathbb{Z}_{N_{01}}$, & $\frac{k}{2\pi} \int_{M^2} dA$ \\
 & $C_{N_0}$ & $\mathbb{Z}_{N_1} \times \mathbb{Z}_{N_2} $ & $k \frac{ N_1 N_2}{(2\pi)^2 N_{012}} \int \o \wedge A^1 \wedge A^2$, $k \in \mathbb{Z}_{N_{012}}$ & $k \frac{ N_1 N_2}{(2\pi)^2 N_{012}} \int_{M^2} A^1 \wedge A^2$ \\
        \specialrule{.5pt}{1pt}{1pt}
3+1d & $C_{N_0}$ & $\mathbb{Z}_{N_1}$ & $k \frac{N_0 N_1}{ (2\pi)^2 N_{01}} \int \o \wedge A \wedge dA$, $k \in \mathbb{Z}_{N_{01}}$ & $k \frac{ N_1}{ (2\pi)^2 N_{01}} \int _{M^3} A \wedge dA$\\
 & $C_{N_0}$ & $\mathbb{Z}_{N_1}$ & $k \frac{N_0 N_1}{(2\pi)^2 N_{01}} \int A \wedge \o \wedge d \o$, $k \in \mathbb{Z}_{N_{01}}$ & $k \frac{ N_1}{2\pi^2 N_{01}} \int_{M^3} A \wedge d\o$ \\
 & $C_{N_0}$ & $\mathbb{Z}_{N_1} \times U(1)$ & $k \frac{N_0 N_1}{(2\pi)^2 N_{01}} \int \o \wedge A^1 \wedge dA^2$, $k \in \mathbb{Z}_{N_{01}}$ & $k \frac{N_1}{(2\pi)^2 N_{01}} \int_{M^3} A^1 \wedge dA^2$ \\
 & $SO(2)$ & $\mathbb{Z}_{N_1} \times \mathbb{Z}_{N_2}$ & $k \frac{N_1 N_2}{(2\pi)^2 N_{12}} \int A^1 \wedge A^2 \wedge d \o$, $k \in \mathbb{Z}_{N_{12}}$ & $k \frac{N_1 N_2}{(2\pi)^2 N_{12}} \int_{M^3} d (A^1 \wedge A^2)$ \\
 & $C_{N_0}$ & $\mathbb{Z}_{N_1} \times \mathbb{Z}_{N_2}$ & $k \frac{N_0 N_1}{(2\pi)^2 N_{01}} \int \o \wedge A^1 \wedge dA^2$, $k \in \mathbb{Z}_{N_{012}}$ & $k \frac{N_1}{(2\pi)^2 N_{01}} \int_{M^3} A^1 \wedge dA^2$ \\
 & $C_{N_0}$ & $\mathbb{Z}_{N_1} \times \mathbb{Z}_{N_2}$ & $k \frac{N_0 N_2}{(2\pi)^2 N_{02}} \int \o \wedge A^2 \wedge dA^1$, $k \in \mathbb{Z}_{N_{012}}$ & $k \frac{ N_2}{(2\pi)^2 N_{02}} \int_{M^3} A^2 \wedge dA^1$ \\
 & $C_{N_0}$ & $\mathbb{Z}_{N_1} \times \mathbb{Z}_{N_2} \times \mathbb{Z}_{N_3}$ & $k \frac{N_0 N_1 N_2 N_3}{(2\pi)^3 N_{0123}} \int \o \wedge A^1 \wedge A^2 \wedge A^3$, $k \in \mathbb{Z}_{N_{0123}}$ & $k \frac{ N_1 N_2 N_3}{(2\pi)^3 N_{0123}} \int_{M^3} A^1 \wedge A^2 \wedge A^3$ \\
         \specialrule{.5pt}{1pt}{1pt}
\begin{minipage}{0.1\textwidth}  3+1d ($*$)\end{minipage} & $SO(2)$ & $U(1)$ & $\frac{k}{2\pi} \int \o \wedge dB$, $k \in \mathbb{Z}$ & $\frac{k}{2\pi} \int_{M^3} dB$ \\
 & $C_{N_0}$ & $Z_{N_1}$ & $\frac{k}{2\pi} \int \o \wedge dB$, $k \in \mathbb{Z}_{N_{01}}$ & $\frac{k}{2\pi} \int_{M^3} dB$ \\
 & $C_{N_0}$ & $\mathbb{Z}_{N_1} \times \mathbb{Z}_{N_2}$ & $k \frac{N_0 N_1 N_2}{(2\pi)^2 N_{012}} \int \o \wedge A \wedge B$ \cite{ye17b}, $k \in \mathbb{Z}_{N_{012}}$ & $k \frac{N_1 N_2}{(2\pi)^2 N_{012}} \int_{M^3} A \wedge B$ \\
        \specialrule{1.5pt}{1pt}{1pt}
\end{tabular}
\caption{\textbf{Generalized Wen-Zee terms as response actions of SPT phases jointly protected by both spatial and internal symmetries}. Here $A^I$ are different flavors of U(1) or $\mathbb{Z}_N$ 1-form gauge fields and $\o$ is the SO(2) spin connection, which is an effective U(1) or $\mathbb{Z}_N$ 1-form gauge field. $N_{ij\dots k} \equiv \text{GCD}\{N_i,N_j,\dots,N_k\}$ where $GCD$ stands for ``greatest common divisor''. The coefficient $k$ or $k$ of each topological response action is quantized to be integral, giving the classification of the corresponding topological phase. It is taken from the finite group in the irreducible classification. For real crystalline materials, the point group is restricted to $C_2,C_3,C_4$ and $C_6$ due to lattice structures in two spacial dimensions. In our discussion of classifications, we relax it to arbitrary finite group, assuming that this finite group symmetry is reduced from some continuum medium. Classifications of SPT phases protected by both spatial and internal symmetries in 3+1d (\textit{denoted by *}), with higher form internal symmetries \cite{Gaiotto2015GGS}  are also presented. For higher form symmetries, the corresponding gauge connections are higher form gauge fields rather than 1-form fields. Here $A, B$ are U(1) or $\mathbb{Z}_N$ 1- and 2-form gauge fields, respectively. In the main text, we omit the wedge product symbol in order to simplify the notation. ``\textit{Irreducible}'' means that any nontrivial subgroup of the total symmetry group would contribute to protecting the SPT phases.  Otherwise, it is said to be ``\textit{reducible}''. The $BF$ terms are implict. For instance, for $G_s = C_{N_0}$ and $G_i = \mathbb{Z}_{N_1} \times \mathbb{Z}_{N_2}$, there are extra terms $\frac{1}{2\pi} \left( N_0 B^0\wedge   d\o + \sum_{I=1}^2 N_I B^I \wedge dA^I \right)$. The angular momentum is the response charge to the $\o$ field defined by $\m{J} = \frac{1}{N_0}\int d^Dx \ \frac{\d S}{\d \o_0}$ for $G_s=C_{N_0}$ and  $\m{J} = \int d^Dx \ \frac{\d S}{\d \o_0}$ for $G_s=SO(2)$, where $D$ is the spatial dimension.
}
\label{tab:Mixed1}
\end{table*}
During our discussion, we assume that $C_{N_0}$ can be any finite group. In real crystalline materials, $C_{N_0}$ is the 2d point group where $N_0=2,3,4,6$, which is restricted by lattice structure.  
Apparently the wedge product forms of these actions look like the twisted terms studied in literatures where only usual gauge fields are involved \cite{Gu2016SPTBF,corbodism3,Ye20163dbSPT,2016arXiv161209298P,Tiwari:2016aa,He2017gaugedSPT,2018arXiv180105416W,Ye20163danomalous,ye16_set,2018arXiv180101638N,PhysRevB.92.045101,WenHeTiwariZhengYe20184dEE}. The unique feature of these topological terms is that the spin connection and usual gauge fields are mixed together to form topological terms, leading to a complex generalization of the  Wen-Zee term previously proposed in the field of the fractional quantum Hall effect \cite{Wen1992WZ}.
 
Inequivalent topological actions can be constructed in the similar way that is utilized in (2+1)d theories. In (2+1)d, with only one gauge field $A$, there exists the standard Chern-Simons (CS) term, $S_1 = \frac{1}{4\pi} \int A\wedge  dA$. With two independent gauge fields $A^1$ and $A^2$, in addition to the self CS terms, there is also the mutual CS term (i.e, the so-called $BF$ term), $S_2 = \frac{1}{2\pi} \int A^1 \wedge   dA^2$. With three or more independent gauge fields $A^i, i=1,2,3,\dots$, there are terms beyond CS terms, $S_3 = \frac{1}{(2\pi)^2} \int A^1 \wedge A^2 \wedge   A^3$, etc. These three types of topological terms are often called Type I,II and III Dijkgraaf-Witten  theories, as the subscripts of the topological terms suggest, because they correspond to inequivalent group cohomologies in Dijkgraaf-Witten  theories.~\cite{Propitius1995DWtheory} If we consider $SO(2)$ connection $\o$, then one of the gauge field should be replaced by $\o$. For instance, $S_3 = \frac{1}{(2\pi)^2} \int \o \wedge A^1 \wedge  A^2$. In (3+1)d, topological terms can be constructed in the similar manner, like $S = \frac{1}{(2\pi)^3} \int \o \wedge A^1 \wedge  A^2 \wedge A^3$ etc. If the gauge groups are finite, there should be some quantized coefficients in front of these topological terms, like what is shown in Table~\ref{tab:Mixed1}. Since classifications of SPT phases are captured by the topological part of response actions, we ignore the non-topological part, like Maxwell terms, in the following discussion.
  
In these theories, $\o$ can be regarded as a gauge field, whose flux is the Riemann curvature flux represented by the Frank vector in a solid. Here we assume that $A^i$ are $\mathbb{Z}_{N_i}$ gauge fields and $\o$ is a $\mathbb{Z}_{N_0}$ gauge field. To discuss the classification of this theory, we need two requirements: a) the large gauge transformation $\D A^i$ and $\D \o$ have periods $2\pi$:
\begin{align}
\oint \D A^i &= 0 \quad (\text{mod } 2\pi), \quad \oint \D \o = 0 \quad (\text{mod } 2\pi); \label{eqn:LGT1}
\end{align}
b) the background fields have $\mathbb{Z}_N$ gauge symmetry:
\begin{align}
\oint A^i &= \frac{2\pi}{N_i} \quad (\text{mod } 2\pi), \quad \oint \o = \frac{2\pi}{N_0} \quad (\text{mod } 2\pi). \label{eqn:ZNsym1}
\end{align}
We also impose Dirac (Monopole) quantization condition
\begin{align}
\int_{M^2} dA^i &= 2\pi n^i, \quad \int_{M^2} d \o = 2\pi n^0. \label{eqn:Diracquant}
\end{align}
Based on Eq.~(\ref{eqn:LGT1}), (\ref{eqn:ZNsym1}) and  (\ref{eqn:Diracquant}), we obtain the classification of inequivalent topological response actions in both 2+1d and 3+1d systems with different internal and spacetime symmetry groups. The result is shown in Table \ref{tab:Mixed1}. Detailed calculations are relegated to Appendix \ref{appendix:CalDetail}. From the third and fourth lines of Table \ref{tab:Mixed1}, for instance, we observe that once we remove the spatial symmetry group, the classifications become trivial, since $\mathbb{Z}_N$ SPT phases in 3+1d have trivial classification (ie., $\mathbb{Z}_1$, see Ref.~\onlinecite{Chen2013CGLWbSPT}). Once we \textit{enrich} the system with spatial symmetry $C_{N}$, we obtain nontrivial SPT phases.

\underline{Irreducible v.s. reducible}. Careful readers may notice  in Table \ref{tab:Mixed1}, there is no discussion on $\int \o \wedge A^2 \wedge  dA^1$ term. The reason is that this term is not a linearly independent term. It can be obtained from the addition of $\int \o \wedge  A^1 \wedge  dA^2$ and $\int A^1 \wedge  A^2 \wedge  d \o$, up to a total derivative term $\int d(\o \wedge  A^1 \wedge  A^2)$. Here is a reminder that in Table~\ref{tab:Mixed1}, we only consider {\it irreducible} classifications, meaning that any nontrivial subgroup of the total symmetry group contributes to protecting the corresponding SPT phases. For instance, for $G_{tot} = C_{N_0} \times \mathbb{Z}_{N_1} \times \mathbb{Z}_{N_2}$, $\int A^1 \wedge A^2 \wedge  dA^2$ also represents a nontrivial SPT phase, but it is {\it reducible} in our sense because $C_{N_0}$ doesn't play any role in this SPT phase.

\subsection{Illustration of topological response phenomena}
\label{sec:Interpretation}
Below we will take some typical examples of generalized Wen-Zee terms and discuss topological response phenomena represented by these topological terms.  

\begin{figure}[htbp]
\includegraphics[width=0.5\textwidth]{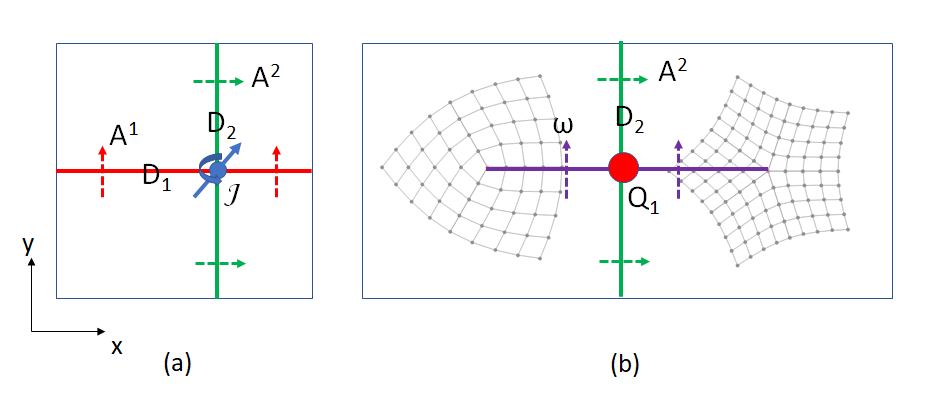}
\caption{(Color online) (a).~Topological response for Eq.~(\ref{eqn:Charge1}). The intersection of $\mathbb{Z}_{N_1}$ and $\mathbb{Z}_{N_2}$ symmetry domain walls $D_1$ and $D_2$ carries the angular momentum $\m{J}$. $A^1$ and $A^2$ are the gauge connections normal to the domain walls, whose integral reflects the angle differences between the two sides of domain walls.  (b).~Topological response of Eq.~(\ref{eqn:Charge1(b)}). The intersection of disclination line and $\mathbb{Z}_{N_2}$ symmetry domain walls $D_2$ carries the $A^1$ charge $\mathcal{Q}_1$. $\o$ and $A^2$ are the gauge connections normal to the disclination line and domain wall, respectively.}
\label{fig:topocharge2D}
\end{figure}

\textbf{\textit{Example 1~~$G=C_{N_0} \times \mathbb{Z}_{N_1} \times \mathbb{Z}_{N_2}$}}~---~ We start from 2+1d as an example. First,~let's consider $S=~k \frac{N_0 N_1 N_2}{(2\pi)^2 N_{012}} \int \o \wedge  A^1\wedge  A^2$.  $N_{ij\dots k} \equiv \text{GCD}\{N_i,N_j,\dots,N_k\}$ where $GCD$ stands for ``greatest common divisor''. Each 1-form gauge connection is defined on the corresponding symmetry domain wall along its the normal direction. Inequivalent topological response theories are labeled by $k\in \mathbb{Z}_{N_{012}}$. The response to the zero component of $\o$, i.e., $\omega_0$, gives us angular momentum:
\begin{align}
\mathcal{J} =& \frac{1}{N_0}\int_{M^2} d^2x \frac{\d S}{\d \o_0} \nonumber\\
=& k \frac{N_1 N_2}{(2\pi)^2 N_{012}} \int_{M^2} d^2x \ \epsilon^{ij} A^1_i A^2_j, \label{eqn:Charge1}
\end{align}
This angular momentum is exactly carried by the gapless energy mode localized at the intersection of $D_1$ and $D_2$  shown in Fig.~\ref{fig:topocharge2D}(a). 
 Furthermore, the angular momentum $\mathcal{J}$ is fractionalized at the minimal (nonzero) value:
\begin{align}
\mathcal{J}_{\text{min}} &= k \frac{N_1 N_2}{(2\pi)^2 N_{012}} \frac{2\pi}{N_1}\frac{2\pi}{N_2}=\frac{k}{N_{012}}
\end{align}
for a given $k$. Since the integer $k$ with identification $k\sim k+N_{012}$ labels distinct SPT phases classified by $\mathbb{Z}_{N_{012}}$, the minimal angular momentum shown above can also be used to detect distinct SPT phases. Let's make the discussion more explicitly. Suppose $N_0=4, N_1=N_2=8$, i.e., $G_s=C_4$ and $G_i=\Z_8\times\Z_8$. Then, $k\sim k+4$ and the minimal nonzero angular momentum for each $k$ is given by (the Planck constant is recovered for the time being):
\begin{align}
\mathcal{J}_{\text{min}}&=\hbar\,, (k=0\,,\text{i.e., the trivial phase})\,;\nonumber\\
\mathcal{J}_{\text{min}}&=\frac{1}{4}\hbar\,, \,\,(k=1);\,\nonumber\\
\mathcal{J}_{\text{min}}&=\frac{1}{2}\hbar\,, \,\,(k=2);\,\nonumber\\
\mathcal{J}_{\text{min}}&=\frac{3}{4}\hbar\,, \,\,(k=3)\,.\nonumber
\end{align}
For a given $k$, all other $\mathcal{J}$'s are multiple of $\mathcal{J}_{\text{min}}$, i.e., 
 $\mathcal{J}=0\,,\pm \mathcal{J}_{\text{min}}\,,\pm2 \mathcal{J}_{\text{min}}\,,\pm3 \mathcal{J}_{\text{min}}\,,\cdots\,$.
 
  We would like to stress that only the fractional part of the angular momentum is meaningful. Indeed, if  $k$ is shifted  by $N_{012}$, the angular momentum $\mathcal{J}$ is simply shifted by an integer.  Formally, once we perform  a large gauge transformation on $A^i$, the angular momentum defined in Eq.~(\ref{eqn:Charge1}) is shifted by an integer. This shift can be understood as attaching trivial SPT phases to the system. We will return to this point in Sec.~\ref{append:chargeshift}. 

We can also study the response to an internal gauge field, say $A^1$. The corresponding $A^1$ charge is 
\begin{align}
\mathcal{Q}_1 =& \frac{1}{N_1}\int_{M^2} d^2x \ \frac{\d S}{\d A^1_0}\nonumber\\
 =& -k \frac{N_0  N_2}{(2\pi)^2 N_{012}} \int_{M^2} d^2x \ \epsilon^{ij} \o_i A^2_j, \label{eqn:Charge1(b)}
\end{align}
which is illustrated in Fig.~\ref{fig:topocharge2D}(b). Furthermore, the charge $\mathcal{Q}_1$ is fractionalized with minimal fractional quantum:
\begin{align}
\mathcal{Q}_1=k\frac{N_0 N_2}{(2\pi)^2 N_{012}} \frac{2\pi}{N_0}\frac{2\pi}{N_2}=\frac{k }{N_{012}}\,.
\end{align}

 \begin{figure}[htbp]
\includegraphics[width=0.31\textwidth]{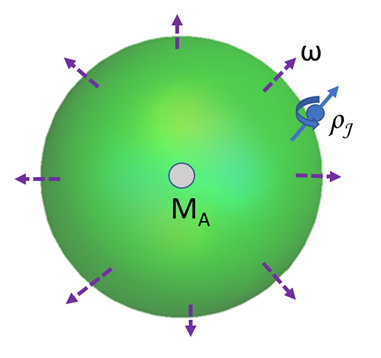}
\caption{(Color online) Topological response for Eq.~(\ref{eqn:Charge2}). $M_A$ is a monopole of gauge field $A$ enclosed by some 2d surface. Since $\o$ and $A$ are U(1) connections, there is a continuous distribution of the angular momentum on the surface where $\o$ is defined along the normal direction of the surface. The integral of angular momentum density ${\m{J}}$ gives Eq.~(\ref{eqn:Charge2}). }
\label{fig:topocharge2D(b)}
\end{figure}
 
\textbf{\textit{ Example 2~~$G = SO(2) \times U(1)$}} --- Another example would be $S = \frac{k}{2\pi} \int \o \wedge  dA$. In this case, in order to have a nontrivial topological action, we must insert monopoles, which is described by Dirac quantization condition (\ref{eqn:Diracquant}). Here we can either insert a U(1) gauge flux, or a curvature flux (disclination) in the real material.  Topological responses give us
\begin{align}
\m{J} &=   \frac{k}{2\pi} \int_{M^2} d^2x \ \epsilon^{ij} \p_i A_j, \label{eqn:Charge2}
\end{align}
or
\begin{align}
\mathcal{Q} &=   \frac{k}{2\pi} \int_{M^2} d^2x \ \epsilon^{ij} \p_i \o_j.
\end{align}
Physically it means that at the defect, the U(1) flux carries the curvature charge, which is the spin or angular momentum; on the other hand, the disclination carries the U(1) charge. In Fig.~\ref{fig:topocharge2D(b)}, it is seen pictorially that once the monopole of U(1) gauge field (spin connection) is enclosed in the curvature flux (U(1) flux) tube, it carries the $\o$ charge (U(1) charge). This phenomenon is essentially Witten's effect.~\cite{Witten1979Witteneffect} Note that both $\o$ and $A$ are continuous gauge connections here. Then there is a continuous distribution of the angular momentum on the surface where $\o$ is defined along the normal direction of the surface in the background of a U(1) monopole of $A$ field. The integral of angular momentum density ${\m{J}}$ gives Eq.~(\ref{eqn:Charge2}). In the language of Ref.~\onlinecite{Wen1992WZ}, due to the mixing term of spin connection and U(1) gauge field, the U(1) flux contributes the spin; on the other hand, the curvature quantum contributes to the electric charge. After integrating over a 2-manifold $M^2$, we have
\begin{align}
N_R &\equiv \int_{M^2} d^2x \ \rho_A = \frac{k}{2\pi} 2\pi n^0 = k n^0,
\end{align}
where $n^0$ is the $\o$ monopole charge. This is related to the shift discussed in Ref.~\onlinecite{Wen1992WZ} in QH systems by a multiplicative factor. 
 
\begin{figure}[htbp]
\includegraphics[width=0.36\textwidth]{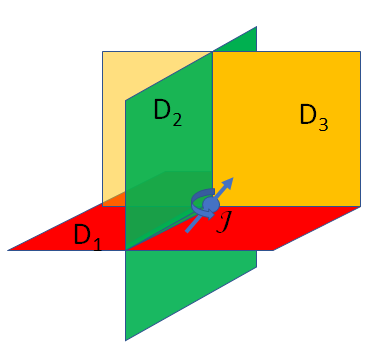}
\caption{(Color online) Topological response phenomena described by Eq.~(\ref{eqn:Charge3}). $D_1$, $D_2$ and $D_3$ are the symmetry $\mathbb{Z}_{N_1}$, $\mathbb{Z}_{N_2}$, $\mathbb{Z}_{N_3}$ domain walls, respectively. The gauge connections $A^1$, $A^2$ and $A_3$ are defined across the corresponding domain walls in the normal directions. The intersection carries the angular momentum of $\o$ labeled by $k \in \mathbb{Z}_{N_{0123}}$. }
\label{fig:topocharge3Da}
\end{figure}

\textbf{\textit{  Example 3~~$G = C_{N_0} \times \mathbb{Z}_{N_1} \times \mathbb{Z}_{N_2} \times \mathbb{Z}_{N_3}$}} --- Now let's move on to the 3+1d system. We first consider $S = k \frac{N_0 N_1 N_2 N_3}{(2\pi)^3 N_{0123}} \int \o \wedge A^1\wedge  A^2\wedge  A^3$. Similar to the 2+1d case, we have the angular momentum: 
\begin{align}
\m{J} &=  k \frac{ N_1 N_2 N_3}{(2\pi)^3 N_{0123}} \int_{M^3} d^3x \ \epsilon^{ijk} A^1_i A^2_j A^3_k\,.\label{eqn:Charge3}
\end{align}
with fractionalized minimal unit $\mathcal{J}_{\text{min}}= \frac{ k}{N_{0123}} $. 
The intersection of the symmetry $\mathbb{Z}_{N_1}$, $\mathbb{Z}_{N_2}$, $\mathbb{Z}_{N_3}$ domain walls $D_1$, $D_2$, $D_3$ carries angular momentum labeled by $k \in \mathbb{Z}_{N_{0123}}$, like Eq.~(\ref{eqn:Charge1}). (c.f.~Fig.~\ref{fig:topocharge3Da})  

\begin{figure}[htbp]
\includegraphics[width=0.46\textwidth]{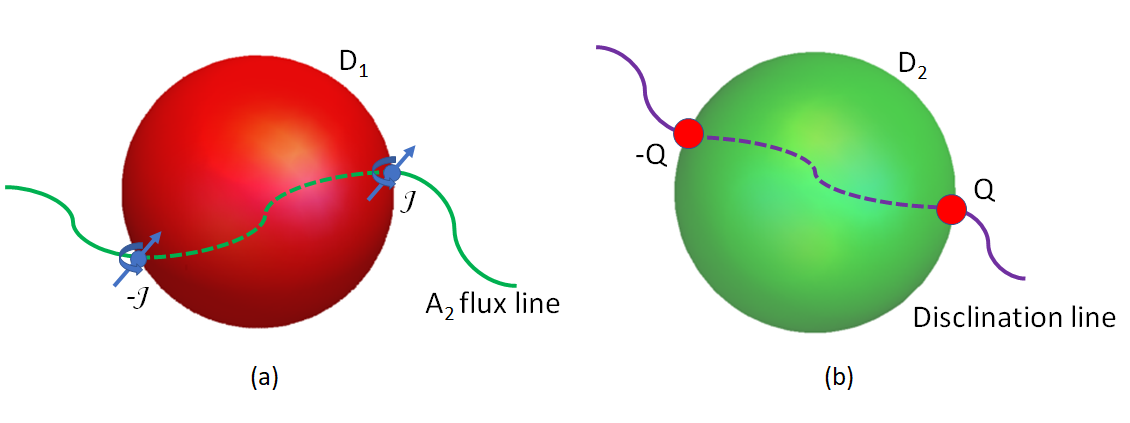}
\caption{(Color online) (a).~Topological response for Eq.~(\ref{eqn:Charge4}).  $D_1$ is the $\mathbb{Z}_{N_1}$ symmetry domain wall. $A^1$ connection is defined across $D_1$ in the normal direction. The intersection of $D_1$ and $A^2$ flux line carries $\o$ angular momentum. (b).~Topological response for Eq.~(\ref{eqn:Charge5}). Similar to (a), the intersections of the disclination line and the $\mathbb{Z}_{N_2}$ symmetry domain wall $D_2$ carry $A^1$ charges. }
\label{fig:topocharge3Db}
\end{figure}

\textbf{\textit{Example 4~~$G=C_{N_0} \times \mathbb{Z}_{N_1} \times U(1)$}} --- For other types of response actions, we have to insert nonzero fluxes to probe the nontrivial SPT order. For instance, from Table \ref{tab:Mixed1}, for $S = k \frac{N_0 N_1}{(2\pi)^2 N_{01}} \int \o \wedge A^1\wedge  dA^2$ in 3+1d, we have
\begin{align}
\m{J} &= k \frac{ N_1}{(2\pi)^2 N_{01}} \int_{M^3} d^3x \ \epsilon^{ijk} A^1_i \p_j A^2_k. \label{eqn:Charge4}
\end{align}
with fractionalized minimal unit $\mathcal{J}_{\text{min}}= \frac{ k}{N_{01}} $. As shown in Fig.~\ref{fig:topocharge3Db}(a), there are angular momenta $\m{J}$ trapped at the intersections between the $\mathbb{Z}_{N_1}$ symmetry domain wall $D_1$ and the $A^2$ flux line.  

\textbf{\textit{ Example 5~~$G=SO(2) \times \mathbb{Z}_{N_1} \times \mathbb{Z}_{N_2}$}}--- Similarly, for $S= k \frac{N_1 N_2}{(2\pi)^2 N_{12}} \int A^1\wedge  A^2 \wedge  d \o$ in 3+1d, 
 we have
\begin{align}
\mathcal{Q}_{1} &= k \frac{ N_2}{(2\pi)^2 N_{12}} \int_{M^3} d^3x \ \epsilon^{ijk} A^2_i \p_j \o_k. \label{eqn:Charge5}
\end{align}
with fractionalized minimal unit $\mathcal{Q}_{\text{min}}= \frac{ k}{N_{12}} $. It means that the intersections between the disclination line and the $\mathbb{Z}_{N_2}$ symmetry domain wall $D_2$ carry  $A^1$ charges, shown in Fig.~\ref{fig:topocharge3Db}(b). 

\textbf{\textit{ Example 6~~$G = SO(2) \times U(1)$} }--- Another interesting example is $S = \frac{k}{2\pi} \int \o \wedge  dB$ in 3+1d, where $B$ is a 2-form gauge field. Analogous to the 2+1d case,
we have
\begin{align}
\m{J} &=  \frac{k}{2\pi} \int_{M^3} d^3x \ \epsilon^{ijk} \p_i B_{jk},
\end{align}
showing that the U(1) flux (i.e. a point in 3d space) carries $\o$ angular momentum. The presence of $B$ field indicates that the corresponding SPT phase has 2-form internal symmetry $G_i$ \cite{Gaiotto2015GGS}. 2-form-symmetry-protected topological phase has been also discussed in Ref.~\cite{ye17b} (see Part 3 of the Supplemental Material of this paper) where  two 1-form and one 2-form internal symmetries are mixed together, resulting in a topological response action like $S\sim \int A^1\wedge A^2\wedge B$ in 3+1d.  The latter, if all gauge fields are dynamical, can support Borromean-Rings statistics among quantized flux loops and gauge charge. The term $B\wedge B$ discussed in Ref.~\cite{bti2} can also be used to describe SPT with two 2-form internal symmetries.

\subsection{Fractionalization of angular momentum  is gauge invariant}
\label{append:chargeshift}
In the above discussions, we have known that the response charge is related to configuration of domain walls and/or disclinations. Below we will explain that only fractional part of response charge / angular momentum is meaningful in characterizing SPT orders. The integer part can be removed by gauge transformations. Let us   perform gauge transformations in two examples. Other topological actions can be analyzed in the similar way. 

{$G = C_{N_0} \times \mathbb{Z}_{N_1} \times \mathbb{Z}_{N_2}$ in 2+1d}-----
The topological term in 2+1d that we are interested in is 
\begin{align}
S &= k \frac{N_0 N_1 N_2}{4\pi^2 N_{012}} \int \o \wedge  A^1\wedge  A^2.
\end{align}
The corresponding $\o$ charge is
\begin{align}
\m{J} &= k \frac{ N_1 N_2}{4\pi^2 N_{012}} \int_{M^2} d^2x A_i^1 A_j^2\epsilon^{ij}. \label{eqn:AppendJ}
\end{align}
Under the gauge transformation
\begin{align}
A^I &\to A^I + d\c^I,\ I=1,2,
\end{align}
we have
\begin{align}
\m{J} &\to k \frac{N_1 N_2}{4\pi^2 N_{012}} \int_{M^2} (A^1+d\c^1)\wedge (A^2+ d\c^2) \nonumber \\
&= \!\m{J}\! + \!k \!\frac{N_1 N_2}{4\pi^2 N_{012}} \!\int_{M^2} \!\!\left(\! d \c^1 \!\wedge \!A^2\! +\! A^1 \!\wedge \!d\c^2 \!+ \!d\c^1\!\wedge \! d\c^2 \!\right) \nonumber \\
&= \!\m{J}\! +\! k \frac{ N_1 N_2}{4\pi^2 N_{012}}\! \bigg[ \!(2\pi m^1)(\frac{2\pi n^2}{N_2}) + (\frac{2\pi n^1}{N_1})(2\pi m^2)    \nonumber \\
& \quad + (2\pi m^1)(2\pi m^2) \bigg] \nonumber \\
&= \m{J} + k \frac{ N_1 N_2}{N_{012}} \left[\frac{m^1 n^2}{N_2} + \frac{n^1 m^2}{N_1} + (m^1 m^2) \right],
\end{align}
where $m^I, n^I \in \mathbb{Z}$. It is easy to see that the extra term is a $k$-dependent integer by noting that $N_{012}$ is the greatest common divisor of $N_0$, $N_1$ and $N_2$. Thus the shift of $\m{J}$ under large gauge transformations can be interpreted as stacking trivial phases in the bulk, and only the fractional part of $\m{J}$ is meaningful.

{$G=C_{N_0} \times \mathbb{Z}_{N_1}$ in 3+1d}----- 
The topological term in 3+1d that we look at is 
\begin{align}
S &= k \frac{N_0 N_1}{4\pi^2 N_{01}} \int \o \wedge A^1 \wedge dA^1.
\end{align}
The corresponding $\o$ charge is
\begin{align}
\m{J} &= k \frac{N_1}{4\pi^2 N_{01}} \int_{M^3} A^1 \wedge dA^1. \label{eqn:AppendJ2}
\end{align}
Under the gauge transformation
 $A^1 \to A^1 + d\c^1$, 
 we have
\begin{align}
\m{J} &\to \m{J} + k \frac{N_1}{4\pi^2 N_{01}} \int_{M^3} d\c^1 \wedge  dA^1 \nonumber \\
&= \m{J} + k \frac{ N_1}{4\pi^2 N_{01}} (2\pi m^1)(2\pi n^1) \nonumber \\
&= \m{J} + k \frac{m^1 n^1  N_1}{N_{01}},
\end{align}
where $m_1,n_1 \in \mathbb{Z}$. We can see that under gauge transformations, $\m{J}$ also shifts by a $k$-dependent integer.

\subsection{Synthetic higher-dimensions}
 
 All topological terms in Table~\ref{tab:Mixed1} are either 2+1d or 3+1d. As a matter of fact, one can easily construct generalized Wen-Zee terms in all other dimensions.  For example, in 1+1d, one can consider $\omega \wedge A$ term. In 4+1d, there are more possible terms, e.g.,   $\omega \wedge A^1\wedge A^2\wedge A^3\wedge A^4$, $\omega \wedge A^1  \wedge A^2\wedge dA^3$, $d\omega \wedge A^1\wedge A^2\wedge A^3$,  $\omega \wedge d A^1\wedge B $, $\omega \wedge  A\wedge C $, where $B$ and $C$ are respectively 2-form and 3-form gauge fields. While 1+1d is physically accessible, 4+1d has also been synthetically realized in experiments by periodically tuning the effective lattice momenta along extra synthetic dimensions \cite{PhysRevLett.111.226401,Lohse:2018aa}. Therefore, even in the condensed-matter framework, it is still physically meaningful to discuss SPTs in higher dimensions. Our bosonic/ spin systems are intrinsically interacting. Theoretical challenge in such interacting systems arises on how to implement interactions in the presence of synthetic dimensions, which was recently discussed in Ref.~\onlinecite{PhysRevX.8.041030}.

\section{Generalized Wen-Zee terms derived from SPT low-energy field theories}
\label{section_bulk_field}
In Sec.~\ref{sec:ClassResponse}, we explored the classification from the topological response actions formed by the generalized Wen-Zee terms in Table~\ref{tab:Mixed1}. One may wonder if there is a bulk field theory description of SPTs such that the above response actions can be derived in a natural way. In this section, we introduce such field theory description in both (2+1) d and (3+1) d and the corresponding gauge transformations. Then,  we introduce a Wen-Zee type of coupling between bulk degrees of freedom and external gauge field as well as spin connection. The latter can be regarded as a background gravity. After bulk degrees of freedom are integrated out, we obtain the generalized Wen-Zee terms discussed in Sec.~\ref{sec:ClassResponse}.  In Appendix~\ref{append:SETaction}, we briefly discuss similar analysis of topological gauge theories for SET phases, gauge transformations and their coupling to background gauge fields. Different from SPT gauge theories, we are not allowed to integrate out the internal gauge fields to obtain the response actions. Otherwise, the coefficient quantization of topological terms after integration will be incorrect. 

 We start from 2+1d systems as an example. The action is given by:
\begin{align}
S =& \int \frac{1}{2\pi} \left(b^0 \wedge d a^0 +\sum^2_{I=1} b^I\wedge   da^I \right)\nonumber\\
& + k \frac{N_0 N_1 N_2}{(2\pi)^2 N_{012}} a^0  \wedge  a^1\wedge   a^2 .
\end{align}
where $a^I,b^I$ are \textit{dynamical} 1-form gauge fields, respectively. This action is invariant under the gauge transformations 
\begin{align}
a^I &\to a^I + df^I, \nonumber \\
b^I &\to b^I + dV^I \nonumber\\
&+ 2\pi k \frac{N_0 N_1 N_2}{(2\pi)^2 N_{012}} \epsilon_{IJK} \left( a^J\wedge  f^K- \frac{1}{2} f^J \wedge  df^K \right),
\end{align}
where $I=0,1,2$, $f^I$ and $V^I$ are zero-form fields. We couple them to the external fields 
\begin{align}
S_b &= -\frac{1}{2\pi} \int \left( \o \wedge  db^0+ \sum^2_{I=1} A^I  \wedge  db^I \right),
\end{align}
where $A^I$ are 1-form external gauge fields. $\omega$ is the spin connection that describes the background gravity induced by the Volterra process in, e.g., Fig.~\ref{fig:Volterra}. We also impose the constraints (\ref{eqn:ZNsym1}).  After integrating out $b^0, b^I$, we obtain $a^0=\o, a^I = A^I$. Then integrating $a^I$ gives us the generalized Wen-Zee term:
\begin{align}
S_{eff} &=  k \frac{N_0 N_1 N_2}{(2\pi)^2 N_{012}} \int \left( \o \wedge   A^1  \wedge  A^2 \right) \label{eqn:3dAction1}
\end{align}
which is the topological response action in the second line of Table \ref{tab:Mixed1}. Similarly, in 3+1d, we have
\begin{align}
S=& \int \frac{1}{2\pi} \left(b^0 \wedge  d a^0 +\sum^3_{I=1} b^I  \wedge  da^I \right) \nonumber\\
&+ k \frac{N_0 N_1 N_2 N_3}{(2\pi)^3 N_{0123}} a^0 \wedge   a^1\wedge    a^2  \wedge  a^3, \label{eqn:4dAction1}
\end{align}
where  $b^I$ are 2-form gauge fields. The action (\ref{eqn:4dAction1}) is invariant under the gauge transformations:
\begin{align}
&a^I \to a^I+df^I, \nonumber \\
&b^I \to b^I + dV^I - \pi k \frac{N_0 N_1 N_2 N_3}{(2\pi)^3 N_{0123}}\epsilon_{IJKL}  \nonumber \\
& \times \left( a^J \wedge  a^K \wedge  f^L - a^J \wedge  f^K \wedge  df^L + \frac{1}{3} f^J \wedge  df^K \wedge  df^L \right)\,,
\end{align}
where $V^I$ now are 1-form fields. After coupling to external fields via 
\begin{align}
S_b &= -\frac{1}{2\pi} \int \left( \o \wedge db^0+ \sum^3_{I=1} A^I\wedge   db^I \right),
\end{align}
and imposing the constraints (\ref{eqn:ZNsym1}), we obtain the generalized Wen-Zee term:
\begin{align}
S_{eff} &= k \frac{N_0 N_1 N_2 N_3}{(2\pi)^3 N_{0123}} \int \o \wedge  A^1\wedge   A^2\wedge  A^3.
\end{align}

\section{Summary and outlook}
\label{sec:Discussion}
In this paper, we have focused on SPTs that are jointly protected by  spatial and internal symmetries. We  have proposed a class of new topological terms---\textit{generalized Wen-Zee terms} and shown how   spatial and internal symmetries are  intertwined. The classification  and response theory of these SPTs has been obtained in the present framework and summarized in Table~\ref{tab:Mixed1}. The physical interpretation of generalized Wen-Zee terms has been analyzed through some typical examples shown in \cref{fig:topocharge2D,fig:topocharge2D(b),fig:topocharge3Da,fig:topocharge3Db}. We have also discussed the relations between these topological response theories and  corresponding bulk low-energy field   theories for SPT  phases. 

Several future directions are summarized below:

\textit{1, Non-Abelian spatial rotation and disclinations.} The
rotational symmetry in the present work is restricted to SO(2)
and its subgroup, i.e. a spatial rotation about a fixed axis.
Nevertheless point group symmetry is not so restricted. A non-
Abelian rotation, e.g., a rotation about one axis followed by
a rotation about another axis. In these situations, disclinations are essentially non-Abelian defects described by Frank
vectors with three non-vanishing components. Probing such
spatial symmetries requires non-Abelian spin connection $\omega_{\mu}^{ij}$. 
Generalizing Wen-Zee term along this line has to study topological terms of non-Abelian fields with a trace operation in SO(3) ``color'' space while the discrete nature of point group can  be guaranteed formally by adding Higgs terms.

\textit{2, Spatial translation  and dislocations.} If the spatial symmetries also involve translation symmetry denoted by $\mathcal{T}$, then the possible expressions of generalized Wen-Zee terms may include  vielbein field in a nontrivial way.  The response phenomenon of each topological term may include an interesting interplay of disclinations, dislocations, and symmetry domain walls.   {For examples, fractionalized momentum may be carried by zero modes that are localized on intersections of symmetry domain walls.} In Ref.~\onlinecite{Barkeshli2012dislocation}, the authors discussed the relation between dislocations and topologies of the manifold in (2+1) d. Specifically, adding a pair of dislocations is equivalent to increasing the genus of the manifold by one. It is interesting to study similar phenomena in higher dimensions and for disclinations, and also recently proposed higher-order SPT phases.~\cite{YouDevakulBurnellNeupert2018HOSPT,SongHuangQiFangHermele2018topologicalcrystal}

\textit{3, Connections to  other approaches.}  Our work on SPTs protected by internal and spatial symmetries focus on the field theoretical aspect. It is desirable to make connections to other approaches to SPT with \textit{at least} spatial symmetries, e.g, real space recipe, method of  abstract algebraic topology \cite{2018arXiv181012308C,SongHuangQiFangHermele2018topologicalcrystal,2018arXiv181011013S,2018arXiv181000801S,2018arXiv181211959G}.

 \textit{4, Canoncal quantization and topological   gravity.} Finally, it is interesting to point out   generalized Wen-Zee terms also represent exotic topological couplings between   gravity and gauge fields with nontrivial non-commuting algebraic structure \cite{Gu2017quantumgravity} 
 once we perform canonical quantization on   generalized Wen-Zee terms.  Throughout this paper, we treat disclinations in materials as static external defects. As a future direction, one can promote these geometric defects to dynamical fields, which corresponds to topological gravity \cite{Dijkgraaf:2018vnm}.

\acknowledgements

We thank Eduardo Fradkin, Xiao-Gang Wen, Yang Qi, Xie Chen, Zheng-Cheng Gu, Shuai Chen, and Shenghan Jiang for helpful communication.  B.H. thanks T.~Li for teaching him Mathematica plotting. This research was supported by the DARPA YFA program, contract D15AP00108 (H.W.), in part by the
National Science Foundation grant DMR 1408713 and DMR 1725401 (P.Y.) at the University
of Illinois, and grant of the Gordon and Betty Moore
Foundation EPiQS Initiative through Grant No. GBMF4305 (P.Y.), by startup grant at Sun Yat-sen University (P.Y.). 

 \appendix
 
%

\section*{Outline}
 {These Appendices are devoted to presenting some technical details for the  results in the main text. While the main text is written in a self-contained and coherent manner, the appendix provides a deeper discussion to details steps. The outline of the appendix is as follows. In Appendix~\ref{appendix_review_WZ}, we briefly review Wen-Zee term in quantum Hall systems. In Appendix~\ref{append:dislocation}, we review the physical interpretation of the dislocation as the torsion flux, in parallel with that of the disclination as the curvature flux. In Appendix~\ref{appendix:CalDetail}, we present the detailed calculations of the classification of SPT phases listed in Table~\ref{tab:Mixed1} in the main text with two examples.  In Appendix~\ref{append:GaussBonnet}, we review the relation between quantization condition for disclinations and Gauss-Bonnet theorem. In Appendix~\ref{append:SETaction}, we also briefly study the bulk field theory of symmetry-enriched topological phases (SETs) with both kinds of symmetries. This logic (i.e., from SPTs to SETs) follows from Ref.~\onlinecite{ye16_set,2018arXiv180101638N} where the field theory of SETs with purely internal symmetries is studied. }

\section{Review on Wen-Zee terms in Quantum Hall systems}\label{appendix_review_WZ}
To be self-contained, we briefly review Wen-Zee term of quantum Hall systems here.~\cite{Wen1992WZ} The long-distance physics of quantum Hall system can be described by the $K$-matrix formalism as
\begin{align}
\m{L} &= \frac{1}{4\pi} \epsilon^{\mu \nu \rho} \left( K_{IJ} a^I_\mu \p_\nu a_\rho + 2t_I A_\mu \p_\nu a^I_\rho + 2s_I \o_\mu \p_\nu a^I_\rho \right).
\end{align}
where $a^I$ are statistical gauge fields, $t_I$ and $s_I$ are the charge and spin vectors, respectively, $A$ is the background electromagnetic U(1) connection and $\o$ is the $SO(2)$ connection.  The term ``$\omega\wedge d a$'' is the original Wen-Zee term. After integrating out the statistical gauge field $a^I$, we have the effective Lagrangian for the background fields
\begin{align}
\m{L}_{eff} &= \frac{1}{4\pi} \epsilon^{\mu \nu \rho} (t_I A_\mu + s_I \o_\mu) \left(K^{-1}\right)^{IJ} \p_\nu (t_J A_\rho + s_J \o_\rho).
\end{align}
The response electromagnetic and spin currents are obtained as
\begin{align}
j_\mu &= \frac{\p \m{L}_{eff}}{\p A^\mu} = \frac{1}{2\pi} \epsilon^{\mu \nu \rho} t_I \left(K^{-1}\right)^{IJ} \p_\nu (t_J A_\rho + s_J \o_\rho), \\
j_{s,\mu} &= \frac{\p \m{L}_{eff}}{\p \o^\mu} = \frac{1}{2\pi} \epsilon^{\mu \nu \rho} s_I \left(K^{-1}\right)^{IJ} \p_\nu (t_J A_\rho + s_J \o_\rho).
\end{align}
Integrating the time components of the above currents, we obtain the electric and spin charges as
\begin{align}
\begin{pmatrix}
N_e \\
N_s
\end{pmatrix} &= \begin{pmatrix}
\mathbf{t}^T {K}^{-1} \mathbf{t} & \mathbf{t}^T {K}^{-1} \mathbf{s} \\
\mathbf{s}^T {K}^{-1} \mathbf{t} & \mathbf{s}^T {K}^{-1} \mathbf{s}
\end{pmatrix} \begin{pmatrix}
N_\phi \\
N_R
\end{pmatrix},
\end{align}
where $N_\phi = \frac{1}{2\pi} \int dA$ and $N_R = \frac{1}{2\pi} \int d\o$ are the flux and curvature quanta, respectively. Then we have
\begin{align}
N_\phi &= \frac{1}{\mathbf{t}^T \mathbf{K}^{-1} \mathbf{t}} N_e - \frac{\mathbf{t}^T {K}^{-1} \mathbf{s}}{\mathbf{t}^T {K}^{-1} \mathbf{t}} N_R.
\end{align}
The first term gives the filling factor $\nu = \mathbf{t}^T {K}^{-1} \mathbf{t}$ and the second term gives the {\it shift} 
\begin{align}
\m{S} &= (\mathbf{t}^T {K}^{-1}\mathbf{s}) \nu^{-1} N_R,
\end{align}
which is a topological quantity, depending on the genus of the manifold but not the metric.

\section{Dislocations as the torsion flux}
\label{append:dislocation}
A dislocation in the media is represented by the Burgers vector $\mathbf{b}=(b^1,b^2,b^3)$, defined in terms of the displacement field as $\oint_C dx^m \p_m u^i(x) = \oint_C dx^m \p_m y^i(x) = -b^i$, $i=1,2,3$, where $C$ is a closed path encircling the dislocation line.~\cite{Katanaev19923dGravityDefect} There is no preferred reference frame in the deformed media when dislocations exist. So we choose an arbitrary curvilinear coordinate system $x^m$, while we use $i,j$ to label the fixed Cartesian reference frame. Although the displacement field is not everywhere smooth in the presence of dislocations, $\p_m y^i$ are everywhere smooth in order to ensure the stress forces  are uniquely defined in every point. Therefore, we can introduce the smooth tensor field $e_m^{\ i}= \p_m y^i$. In the language of geometry, it is the vielbein field. When there are dislocations, the integrability condition breaks down: $\p_m e_n^{\ i} - \p_n e_m^{\ i} \neq 0$. The Burgers vector can be expressed as $\iint_S dx^m dx^n (\p_m e_n^{\ i} - \p_n e_m^{\ i}) = -b^i$. By Stokes' theorem, it can also be written as $\oint_C dx^m e_m^{\ i} = -b^i$. In fact, $e_m^{\ i}$ can be used to describe the distribution of dislocations in the media. Let's introduce the torsion tensor $T_{mn}^{\quad i} = \p_m e_n^{\ i} - \o_m^{\ ij} e_{nj} -(m \leftrightarrow n)$, where $\o_m^{\ ij}$ is the $SO(3)$-connection. In the absence of curvature, we can choose $SO(3)$-connection as a pure gauge and set $\o_m^{\ ij} =0$. Then the Burgers' vector becomes $\iint_S dx^m dx^n T_{mn}^{\quad i} = -b^i$. When the curvature exists, the above equation still holds. Nevertheless, in the continuum media with continuously distributed dislocations, only the local Burgers vector is well defined: $dx^m dx^n T_{mn}^{\quad i} = - d b^i(x)$.~\cite{Katanaev19923dGravityDefect} Similar arguments work for disclinations. From the infinitesimal form, it is apparent that torsion can be interpreted as the surface density of the Burgers vector. By Laughlin argument, an insertion of dislocation induces momentum pumping. In this sense, momentum is the {\it charge} if we regard the vielbein field as a gauge field. 

\section{Derivations of classifications in Table \ref{tab:Mixed1}}
\label{appendix:CalDetail}
We show calculations of the classification for two topological terms as examples. Other topological terms can be analyzed in the same way. From the calculations, we will see that the classification comes from two steps: large gauge transformations and shifts. The large gauge transformation tells us the quantization of the coefficient of the topological action, while the shift tells us the period, thus classification, of the topological action. 

\subsection{$G=C_{N_0} \times \mathbb{Z}_{N_1}$}
The topological action is
\begin{align}
S &= \int \left( \frac{1}{2\pi} (N_0 B^0 d \o + N_1 B^1 dA^1) + p \o A^1 dA^1 \right), \label{T1}
\end{align}
which is invariant under the gauge transformations
\begin{align}
\o &\to \o + d \c^0, \nonumber \\
A^1 &\to A^1 + d \c^1, \\
B^I &\to B^I + dV^I - \frac{2\pi p}{N_I} \epsilon^{IJ} \c^J dA^1, \nonumber
\end{align}
up to total derivative terms, where $\epsilon^{01} = -\epsilon^{10} =1$. Under large gauge transformations, 
\begin{align}
\frac{1}{2\pi} \int_{M^3} dB^0 &\to \frac{1}{2\pi} \int_{M^3} dB^0 - \frac{1}{2\pi} \int_{M^3} \frac{2\pi p}{N_0} d\c^1 dA^1 \nonumber \\
&= \frac{1}{2\pi} \int_{M^3} dB^0 - \frac{p}{N_0} (2\pi m^1) (2\pi n^1), \nonumber \\
\Rightarrow 4\pi^2 p & \in N_0 \mathbb{Z}, \label{LG1}
\end{align}
where $m^1,n^1 \in \mathbb{Z}$. Similarly, 
\begin{align}
\frac{1}{2\pi} \int_{M^3} dB^1 &\to \frac{1}{2\pi} \int_{M^3} dB^1 + \frac{p}{N_2} (2\pi m^0)(2\pi n^2), \nonumber \\
\Rightarrow 4\pi^2 p &\in N_1 \mathbb{Z}. \label{LG2}
\end{align}
From Eq.~(\ref{LG1}),(\ref{LG2}), we have
\begin{align}
4\pi^2 p &= k \text{lcm}(N_1,N_2) \nonumber \\
\Rightarrow  p &= \frac{k N_0 N_1}{4\pi^2 N_{01}}, \quad k \in \mathbb{Z},
\end{align}
where $k \in \mathbb{Z}$. 

Eq.~(\ref{T1}) is invariant under the shift
\begin{align}
dB^0 &\to dB^0 - \frac{k N_1}{2\pi N_{01}} A^1 dA^1, \nonumber \\
dB^1 &\to dB^1 + \frac{K_1 N_0}{2\pi N_{01}} \o dA^1, \\
k &\to k+k+K_1, \nonumber
\end{align}
where $k,K_1 \in \mathbb{Z}$. Then under the shift, 
\begin{align}
\D \left( \frac{1}{2\pi} \int N_0 \o dB^0 \right) &= - \frac{k m^0 m^1 2\pi n^1}{N_{01}} \in 2\pi \mathbb{Z}. \nonumber \\
\Rightarrow k &= \ell N_{01}. \label{S1}
\end{align}
Similarly, change in $\frac{1}{2\pi} \int N_1 A^1 dB^1$ gives us
\begin{align}
K_1 &= \ell' N_{01}. \label{S2}
\end{align}
Combining Eq.~(\ref{S1}) and (\ref{S2}), we have 
\begin{align}
k &\sim k+N_{01},
\end{align}
which is essentially the classification of the topological system described by Eq.(\ref{T1}). 

\subsection{$G=C_{N_0} \times \mathbb{Z}_{N_1} \times \mathbb{Z}_{N_2}$}
The topological action is
\begin{align}
S &= \int \big( \frac{1}{2\pi} (N_0 B^0 d\o + N_1 B^1 dA^1 + N_2 B^2 dA^2) \nonumber \\
& \quad + p \o A^1 dA^2 \big), \label{T2}
\end{align}
which is invariant under the gauge transformations
\begin{align}
\o &\to \o + d\c^0, \nonumber \\
A^I &\to A^I + d\c^I, \quad I=1,2 \nonumber \\
B^I &\to B^I + dV^I - \frac{2\pi p}{N_I} \epsilon^{IJ} \c^J dA^2, \quad I,J=0,1, \nonumber \\
B^2 &\to B^2 + dV^2, 
\end{align}
up to total derivative terms. The large gauge transformations on $\frac{1}{2\pi} \int dB^0$ and $\frac{1}{2\pi} \int dB^1$ give us 
\begin{align}
p &= \frac{k N_0 N_1}{4\pi^2 N_{01}}, \quad k \in \mathbb{Z}.
\end{align}
Eq.~(\ref{T2}) is invariant under the shift
\begin{align}
dB^0 &\to dB^0 + \frac{k N_1}{2\pi N_{01}} A^1 dA^2, \nonumber \\
dB^1 &\to dB^1 - \frac{K_1 N_0}{2\pi N_{01}} \o dA^2, \nonumber \\
B^2 &\to B^2 - \frac{K_2 N_0 N_1}{2\pi N_{01} N_2} \o A^1, \nonumber \\
k &\to k+k +K_1 +K_2.
\end{align}
Under the shift, $\D (\frac{1}{2\pi} \int dB^0)$ and $\D (\frac{1}{2\pi} \int dB^1)$ give us 
\begin{align}
k &= \ell N_{01}, \quad K_1 = \ell' N_{01}.
\end{align}
For $B^2$, we have
\begin{align}
\D(\frac{1}{2\pi} \int dB^2) &= - \frac{K_2 N_0 N_1}{4\pi^2 N_{01} N_2} \int (d\o A^1 - \o dA^2) \nonumber \\
&= - \frac{K_2}{N_2} \frac{N_0 m^1 + N_1 m^0}{N_{01}} \in 2\pi \mathbb{Z}. \nonumber \\
\Rightarrow K^2 &= \ell'' N_2.
\end{align}
Then we have
\begin{align}
k &\sim k+N_{012}.
\end{align}

\section{Dirac quantization condition for disclinations on orientable manifolds and Gauss-Bonnet theorem}
\label{append:GaussBonnet}
As is familiar from QH systems, the curvature flux in 2d orientable manifold $M^2$ is quantized by Gauss-Bonnet theorem
\begin{align}
\chi &= \frac{1}{2\pi} \int_{M^2} d \bar{\o} = \frac{1}{4\pi} \int d^2x \sqrt{h} R = 2(1-g),
\end{align}
where $\chi$ is the Euler characteristic,  $\bar{\o}$ is the spin connection defined in Ref.~\onlinecite{Wen1992WZ}, in dictinction with $\o$ defined in the main text; $h$ is the metric and $R$ is the Ricci scalar of $M^2$. In terms of $\bar{\o}$, we have on sphere $\int_{S^2} d\bar{\o} = 4\pi n$, $n \in \mathbb{Z}$. Note that in Ref.~\onlinecite{Wen1992WZ}, the mininum quantum of angular momentum is 1/2, while here our minimum quantum is 1. Taking this into account, our quantization condition in Eq.~(\ref{eqn:Diracquant}) is consistent with the quantization condition given by Gauss-Bonnet theorem. As a result, our spin connection $\o$ here satisfies the same quantization condition as a normal U(1) gauge connection. In (3+1) d, the spin connection is generically non-Abelian, even for the spatial part only. In this case, it is hard to define the curvature flux. To resolve this problem, we restrict our discussion of the geometric response on a two-dimensional submanifold, namely, we decompose the spacetime manifold $M^4 = M^2 \times M'^2$. Then the curvature flux on the submanifold, say, $M^2$ is well-defined, like the QH systems. Then we put internal gauge fluxes or monopoles on the other submanifold $M'^2$ such that the total topological action is nontrivial. For more general four manifolds, we can choose the vielbein field as \cite{Gromov2016bdyQHE} $e^\mu_{\ a} = (\d^\mu_{\ t}, e^\mu_{\ x}, e^\mu_{\ y}, \d^\mu_{\ z})$ such that only $xy$ plane has nontrivial geometry. 

\section{Topological field theories for SET phases}
\label{append:SETaction}
Similar to the last subsection, we can turn the level of BF terms to be integers larger than one. The resulting theories have intrinsic topological orders and describe SET phases. There is a caveat, though. The path-integral over bulk degrees of freedom can be legitimately performed in SPTs but not in SETs since nontrivial topological excitations in latter lead to inconsistency on level quantization. In this case, we have for 2+1d,
\begin{align}
S &= \int \left( \frac{n_0}{2\pi} b^0 d a^0 + \sum^2_{I=1} \frac{n_I}{2\pi} b^I da^I + k \frac{N_0 N_1 N_2}{(2\pi)^2 N_{012}} a^0 a^1 a^2 \right), \label{eqn:3dSET1}
\end{align}
which is invariant under the gauge transformations up to surface terms
\begin{align}
a^I &\to a^I + df^I, \nonumber \\
b^I &\to b^I + \frac{1}{n_I} dV^I + \frac{2\pi k}{n_I} \frac{N_0 N_1 N_2}{(2\pi)^2 N_{012}} \epsilon^{IJK} (a^J f^K - \frac{1}{2} f^J df^K),
\end{align}
where $I=0,1,2$, $f^I$ and $V^I$ are zero-form fields. We can again introduce couplings to external fields
\begin{align}
S_b &= -\frac{1}{2\pi} \int \left( \o db^0+ \sum^2_{I=1} A^I  db^I \right),
\end{align}
and impose the constraints (\ref{eqn:ZNsym1}). The equations of motion for $b^0,b^I$ are $n_0 d\o + dA^0 = 0, 
n_I da^I + dA^I =0$. One should note that in this case, we can't integrate out the dynamical fields $a^I,b^I$ to obtain an effective response theory since the bulk of the SET phase is nontrivial. There are nontrivial interactions between the external fields $\o,A^I$ and the intrisic dynamical fields. Similarly, in 3+1d, we have 
\begin{align}
S &= \int \left( \frac{n_0}{2\pi} b^0 d a^0 + \sum^3_{I=1} \frac{n_I}{2\pi} b^I da^I + k \frac{N_0 N_1 N_2 N_3}{(2\pi)^3 N_{0123}} a^0 a^1 a^2 a^3 \right), \label{eqn:4dSET1}
\end{align}
which is invariant under the gauge transformations up to surface terms
\begin{align}
a^I &\to a^I+df^I, \nonumber \\
b^I &\to b^I + \frac{1}{n_I} dV^I \nonumber \\
& \quad - \frac{\pi k}{n_I} \frac{N_0 N_1 N_2 N_3}{(2\pi)^3 N_{0123}} \nonumber \\
& \quad \times \epsilon_{IJKL} \left( a^J a^K f^L - a^J f^K df^L + \frac{1}{3} f^J df^K df^L \right).
\end{align}

%


\end{document}